\documentclass[%
 reprint,
 amsmath,amssymb,
 aps,
]{revtex4-1}

\usepackage{graphicx}
\usepackage{dcolumn}
\usepackage{bm}
\usepackage{diagbox}

\begin{document}

\preprint{APS/123-QED}

\title{A single-state semi-quantum key distribution protocol and its security proof }

\author{Wei Zhang$^{1,2,3}$}
 \altaffiliation[]{aandy1224zw@163.com}
\author{Daowen Qiu$^{1,2}$}%
 \email{issqdw@mail.sysu.edu.cn (Corresponding author)}

\affiliation{%
$^{1}$School of Data and Computer Science, Sun Yat-sen University, Guangzhou 510006, China \\
$^{2}$The Guangdong Key Laboratory of Information Security Technology, Sun Yat-sen University, Guangzhou 510006, China\\
$^{3}$School of Mathematics and Statistics, Qiannan Normal College for Nationalities, Duyun 558000, China\\
}%
\date{\today}

\begin{abstract}
Semi-quantum key distribution (SQKD) can share secret keys by using less quantum resource than its fully quantum counterparts, and this likely makes SQKD  become more practical and realizable.  In this paper, we present a new  SQKD protocol by introducing the idea of B92 into semi-quantum key distribution and prove its unconditional security.  In this protocol, the sender Alice just sends one qubit to the classical Bob and Bob just prepares one state in the preparation process.  Indeed the classical user's measurement is not necessary either. This protocol can reduce some quantum communication and make it easier to be implemented. It can be seen as the semi-quantum version of B92 protocol, comparing to the protocol BKM2007 as the semi-quantum version of BB84 in fully quantum cryptography.   We verify it has higher key rate and therefore is more efficient. Specifically we prove it is unconditionally secure by computing a lower bound of the key rate in the asymptotic scenario from information theory aspect. Then we can find a threshold value of errors such that for all error rates less than this value, the secure key can be established between the legitimate users definitely. We make an illustration of how to compute the threshold value in case of the reverse channel is a depolarizing one with parameter $p$. Though the threshold value is a little smaller than those of some existed SQKD protocols, it can be comparable to the B92 protocol in fully quantum cryptography.

\end{abstract}

\pacs{Valid PACS appear here}
\maketitle


\section{Introduction}
Semi-quantum key distribution (SQKD) is a new technique to share secure secret keys in quantum world. In an SQKD, one of the users is restricted to measure, prepare and send qubit in a fixed computational basis. We call it the semi-quantum or classical user. Boyer et al \cite{1} designed the first SQKD protocol to share secret keys between quantum Alice and classical Bob successfully in 2007 (BKM07). In BKM07 protocol, Alice prepares qubits in two different basis randomly and sends them to Bob, and Bob can do two kinds of operations when he receives the state as follows:
\begin {enumerate}
\item  SIFT: Bob chooses to measure the qubit and resend a new one to Alice. He measures the state he received in the computational basis $Z=\{|0\rangle, |1\rangle\}$ and resends the result state to Alice. In other words, Bob sends the state $|i\rangle(i\in\{0,1\})$ to Alice when he gets the measurement outcome $i$.

\item  CTRL: Bob chooses to reflect it back. He just makes the state pass through and returns it to Alice. Under this circumstance, Bob knows nothing about the transit qubit because he cannot gain any information.
\end {enumerate}
When Alice gets the returning state, she measures it in the $Z$-basis or $X$-basis randomly. When Bob chooses to SIFT and Alice chooses to measure in the $Z$-basis, they share a bit.

From the above scheme, we can see that Bob just does some classical performances, making the SQKD protocol more practical and realizable. Since the first SQKD was proposed by Boyer  et al.  \cite{1}, various SQKD protocols have been provided \cite{2,3,4,5,6,7,8,9,10,11,12}. Specifically,  Zou and Qiu et al. \cite{4} proposed five SQKD protocols with less than four quantum states based on BKM07 and proved them to be completely robust. A multi-user protocol was developed in Ref. \cite{5},  establishing secret keys between a quantum user and several classical ones. In Ref. \cite{6}, an SQKD protocol was proposed based on quantum entanglement. Zou and Qiu et al. \cite{10} presented an SQKD protocol without invoking the classical party's measurement ability. Krawec \cite{9} designed a mediated SQKD protocol allowing two semi-quantum users to share secure secret keys with the help of a quantum server. Recently, Krawec \cite{12} has proposed a  single-state SQKD protocol, in which the classical Bob's reflection can contribute to the raw key.

 SQKD protocols mainly rely on a two-way quantum channel, which leads to the eavesdropper Eve having two opportunities to attack the transit qubits during their transmission. This may increase the possibility for Eve to gain more information on $A$ or $B$'s raw key and make the security analysis more complicated.  Most of the existing SQKD protocols are limited to discuss their robustness rather than unconditional security. A protocol is said to be robust if any attacker can get nontrivial information on  $A$ or $B$'s secret key, the legitimate users can detect his existing with nonzero probability \cite{3}. Then the robustness of SQKD protocols can only assure any attack can be detected, but it cannot tell us how much noise the protocol can tolerate to distill a secure key after applying the technique of error correction and privacy amplification.

Recently, the situation has been improved. In Ref. \cite{13}, the relationship between the disturbance and the amount of information gained by Eve was provided under the circumstance that Eve just performs individual attacks. Krawec \cite{14} proved that any attack operator was equivalent to a restricted attack in a single-state SQKD protocol. Then Krawec \cite{15} further proved the unconditional security of BKM2007 by giving the lower bound on the key rate in the asymptotic scenario. To the best of my knowledge, this is the first unconditional security proof of an SQKD protocol. Furthermore, Krawec \cite{16} proved the unconditional security of a mediated SQKD protocol allowing two semi-quantum users to share secure secret key with the help of a quantum server even under the circumstance that the quantum server is an all-powerful adversary. Recently, Krawec\ \cite{12} has provided an unconditional security proof of a single-state SQKD protocol.

In this paper, we introduce the idea of $B92$ into semi-quantum key distribution and design a new single-state SQKD protocol. It can be seen as the semi-quantum version of B92, comparing to the Protocol BKM07 as the semi-quantum version of BB84 in fully quantum cryptography. Additionally, the classical Bob has no need to own measurement equipment and the CTRL-bit can contribute to the raw key, in other words, the classical Bob's reflection can contribute to the raw key. All of these make our protocol to be more practical and efficient. In addition, we prove it to be unconditional secure by finding a threshold value such that all the error rates less than this value, the secure keys can be established definitely.

The rest of this paper is organized as follows. First, in Section 2 we give some preliminaries. Then in Section 3 we present our single-state SQKD protocol. In particular, in Section 4 we give the unconditional security proof in detail. Finally in section 5 we  make a short conclusion and give some issues for  future consideration.

\section{Preliminaries}
In this section, we give some preliminaries and some notations which are about to appear in the next sections.

The computational basis denoted as $Z$ basis is the two state set $\{|0\rangle, |1\rangle\}$, the $Hadamard$ basis denoted as $X$ basis is the set $\{|+\rangle, |-\rangle\}$, where
\begin{eqnarray}
&& |+\rangle = \frac{|0\rangle+|1\rangle}{\sqrt{2}},\\
&& |-\rangle = \frac{|0\rangle-|1\rangle}{\sqrt{2}}.
\end{eqnarray}

Given a complex number $z\in\mathbb{C}$, we denote $Re(z)$ and $Im(z)$ as its real and imaginary components respectively. The conjugate of $z$ is denoted as $z^{*}$.
If $U$ is a complex matrix (operator), its conjugate transpose (conjugate) is denoted as $U^{\dag}$.

Consider a random variable $X$. Suppose each realization $x$ of $X$ belongs to the set $N=\{1,2,\cdots,i,\cdots,n\}$. Let $P_{X}(i)$ be the probability distribution of $X$. Then the Shannon entropy of $X$ is
\begin{eqnarray}
&& H(X)=H(P_{X}(1),\cdots,P_{X}(i),\cdots,P_{X}(n))\\\nonumber
&& \quad \quad \quad =-\sum^{n}_{i=1}P_{X}(i)\log_{2}(P_{X}(i)).
\end{eqnarray}
Note that here we define $0\log_{2}0=0$. When $N=2$, $H(X)=h(P_{X}(1))$, where $h(x)=H(x,1-x)$ is the Shannon binary entropy function.

Let $\rho$ be a density operator acting on an $n$-dimensional Hilbert space $\mathcal{H}$ satisfying
\begin{eqnarray}
&& \rho =\sum^{n}_{i=1}\lambda_{i}|i\rangle\langle i|,
\end{eqnarray}
where $\lambda_{i}(i=1,2,\cdots,n$) is the $i$-th eigenvalue of $\rho$ and $\{|1\rangle, |2\rangle,\cdots, |n\rangle\}$ is the standard basis of $\mathcal{H}$. Then we denote $S(\rho)$ as its von Neumann entropy such that
\begin{eqnarray}
&& S(\rho) =H(\{\lambda_{i}\}_{i})=-\sum^{n}_{i=1}\lambda_{i}\log_{2}\lambda_{i}.
\end{eqnarray}

Let $\rho$ be a classical quantum  state expressed as
\begin{eqnarray}
&& \rho =\sum^{n}_{i=1}P_{X}(i)|i\rangle\langle i|\otimes \rho_{i}.
\end{eqnarray}
Then
\begin{eqnarray}
&& S(\rho) =H(P_{X}(i))+\sum_{i=1}^{n} P_{X}(i)S(\rho_{i}) .
\end{eqnarray}

If $\rho_{AB}$ is a density operator acting on the bipartite space $\mathcal{H}_{A}\otimes \mathcal{H}_{B}$, we use $S(AB)$ to denote the von Neumann entropy of $\rho_{AB}$ and $S(B)$ the von Neumann entropy of $\rho_{B}$ where $S(B)=S(tr_{A}(\rho_{AB}))$. We use $S(A|B)$ to denote the von Neumann entropy of $A$'s system conditioned by system $B$ such that
\begin{eqnarray}
S(A|B) =S(AB)-S(B)=S(\rho_{AB})-S(tr_{A}\rho_{AB}).
\end{eqnarray}

Let $N$ be the size of $A$ and $B$'s raw key of an SQKD protocol, and $\ell(N)< N$ denotes the size of secure secret key distilled after error correction and privacy amplification. Let $r$ denote the key rate in the asymptotic scenario ($N \rightarrow \infty$). Then
\begin{eqnarray}
&& r=\lim_{N\rightarrow \infty}\frac{\ell(N)}{N}\geq \inf (S(B|E)-H(B|A)),
\end{eqnarray}
where $H(B|A)$ is the conditional Shannon entropy and the infimum is over all attack strategies an Eve can perform \cite{12,17,18}.

\section{The protocol}
In this section, we present our single-state SQKD protocol, in which the receiver Bob is limited to be classical. The protocol consists of the following steps:

\begin{enumerate}
\item Alice prepares and sends $N$ quantum states $|+\rangle$ to Bob one by one, where $N=4n(1+\delta)$, $n$ is the desired length of the INFO string, and $\delta > 0$ is a fixed parameter. Alice sends a quantum state only after receiving the previous one.
\item Bob prepares $\frac{N}{2}$ quantum states $|0\rangle$ and generates a random string $\hat{K}_{B}\in\{0,1\}^{N}$ to be his candidate raw key. Bob chooses SIFT or CTRL randomly. Here CTRL means reflecting it back with no disturbance and SIFT means discarding the state he received and sending $|0\rangle$ to Alice instead.

(1) Define $\hat{K}^{(i)}_{B}=0$, when Bob chooses CTRL.

(2) Define $\hat{K}^{(i)}_{B}=1$, when Alice chooses SIFT.

\item Alice also generates a random string $\hat{K}_{A}\in \{0,1,-1\}^{N}$ to be her candidate raw key. When she measures the $i$-th quantum state in the $Z$ basis and gets the outcome $1$, she sets $\hat{K}^{(i)}_{A}=0$. When she measures the $i$-th quantum state in the $X$ basis and gets the outcome $-$, she sets $\hat{K}^{(i)}_{A}=1$. Otherwise, she sets $\hat{K}^{(i)}_{A}=-1$. Then we can get $P(\hat{K}^{(i)}_{A}=-1)=\frac{1}{2}$, where $P(x)$ denotes the probability of $x$.
\item Alice announces Bob to drop all the iterations when $\hat{K}^{(i)}_{A}=-1$ through authenticated classical channel shared previously. Then Alice and Bob will get $K_{A}, K_{B}\in\{0,1\}^{l}$ to be their raw key respectively. Then $l$ is expected to approximate $\frac{N}{2}$. They abort the protocol when $l<2n$.

\item Bob chooses at random $n$ bits from his raw key $K_{B}$ to be TEST bits and announces their positions and values respectively by the authenticated classical channel. Alice checks the error rate on the TEST bits. If it is higher than some predefined threshold value $P_{T}$, they abort the protocol.

\item Alice and Bob select the first $n$ remaining bits of $K_{A}$ and $K_{B}$ respectively to be their INFO string.

\item Alice announces ECC (error correction code) and PA (privacy amplification) data, she and Bob use them to extract the $m$-bit final key from the $n$-bit INFO string.
\end{enumerate}

Note that, we can make the classical Bob to prepare qubit $|1\rangle$ instead of $|0\rangle$ when he SIFTs the qubit. Correspondingly, Alice should set $\hat{K}^{i}_{A}=0$ when she measures in $Z$ basis and gets measurement outcome $0$.

Next, we prove our protocol is correct. Assume $K^{i}_{A}=0$, according to the protocol, we will conclude that Alice performs measurement in the computational basis and gets the outcome $1$. Then we can infer that the qubit she received is bound to be $|+\rangle$ if there is no disturbance. Therefore, Bob's raw key bit $K^{i}_{B}$ should be $0$ definitely. When $K^{i}_{A}=1$, Alice measures in the $X$ basis and gets the result $-$. Then we can infer Alice's receiving qubit is $|0\rangle$ definitely. Consequently, $K^{i}_{B}=1$. From the above protocol, we can see Alice's raw key bit $K^{i}_{A}$ is perfectly correlated to Bob's raw key bit $K^{i}_{B}$ in each iteration in case of no disturbance exists. Then we can conclude that our protocol is correct.

In order to illustrate a protocol's efficiency uniformly, we define a parameter $\ell=\lim_{n\rightarrow\infty}\frac{n}{N}$ , where $n$ is the length of INFO string and $N$ is the number of qubit transmitted in the quantum channel, including the forward and reverse channel. Then we can get our protocol's efficiency parameter $\ell=\frac{1}{8}$.

Compared with the single-state SQKD protocol in  \cite{4}, the classical Bob's measurement equipment can be removed, which makes our protocol is easier to implemented. Besides, the CTRL bits can contribute to the raw key, which makes our protocol  get higher key rate to be more efficient. Specifically, the efficiency parameter $\ell$ of protocol in  \cite{4} is $\frac{1}{16}$.

In comparison with the protocol in \cite{10}, Alice just sends one qubit to Bob and Bob just prepares one state in the preparation process in each iteration, which makes our protocol be able to reduce some quantum communications. Additionally, our protocol is more efficient because the CTRL bits can contribute to the raw key. The efficiency parameter $\ell$ of the protocol in \cite{10} is less than $\frac{1}{12}$.

With respect to Krawec's newly protocol in \cite{12}, the classical Bob can be further restricted to have no measurement ability and he just prepares one state when choosing to SIFT in our protocol. In addition, some iterations have to be discarded to balance the probability of Bob's raw key bits in Krawec's protocol which makes it less efficient inevitably.

In order to make a clear illustration, we using TABLE I to demonstrate the main advantages compared to some existed semi-quantum key distribution protocols as follows:

\begin{table}[h]
\center
\begin{tabular}{|l|c|c|c|c|c|}
 \hline
 \diagbox{$i$}{$i'$} &$1'$ & \quad $2'$ \quad & $3'$ & \quad $4'$ \quad  &$5'$  \\
 \hline
  $1$ & $\{|0\rangle, |1\rangle,|+\rangle,|-\rangle\}$ &  Y  & $\{|0\rangle, |1\rangle\}$ &  N & $\frac{1}{16}$
\\
\hline
$2$  & $|+\rangle$ & Y &$\{|0\rangle, |1\rangle\}$ & N & $\frac{1}{16}$  \\
 \hline

$3$  & $\{|0\rangle, |1\rangle,|+\rangle,|-\rangle\}$& N & $\{|0\rangle, |1\rangle\}$ & N & less than $\frac{1}{12}$  \\
 \hline

$4$  & $|+\rangle$ & Y & $\{|0\rangle, |1\rangle\}$ & Y & $\frac{3}{28}$   \\
 \hline

$5$  & $|+\rangle$ & N& $\{|0\rangle\}$ & Y & $\frac{1}{8}$   \\
 \hline
 \end{tabular}
 \caption{ Comparison of our protocol with some existed ones. Notations: 1 protocol BKM07; 2 Zou's single-state protocol in Ref \cite{4}; 3 Zou's protocol in Ref \cite{10}; 4 Krawec's single-state protocol in Ref \cite{12}; 5 our single-state protocol; $1'$ quantum states prepared by the sender; $2'$ the classical receiver measures the received qubits or not; $3'$ quantum states generated by the receiver; $4'$ CTRL-bit contributes to the raw key or not; $5'$ the efficiency parameter $\ell$.}
 \center
 \end{table}

 From TABLE I, we can see our single-state SQKD protocol is not only more efficient but also easier to implement. Next, we show it is also unconditionally secure.

\section{Security proof}

Firstly, we restrict our security proof on Eve's collective attack. Then we spread it into the circumstance of general attack. Collective attack is a typical attack strategy that Eve performs the same operation in each iteration of the protocol and postpones to measure her ancilla until any future time. General attack (coherent attack or joint attack) is a kind of more powerful attack that Eve can perform any operation allowed by the laws of quantum physics and postpone her measurement all by herself \cite{20}.

\subsection{Modeling the protocol}

We use $\mathcal{H}_{A}$, $\mathcal{H}_{B}$ and $\mathcal{H}_{E}$ to denote Alice, Bob and Eve's Hilbert spaces respectively. $\mathcal{H}_{T}$ is the Hilbert space of the transit states. Generally, they are all assumed to be finite. In order to make a clear illustration, we just take one iteration for example to prove the unconditional security.

Krawec \cite{14} has pointed out any collective attack $(U_{F}, U_{R})$ is equivalent to a restricted operation $(b, U)$ where $b\in (-\frac{1}{2},\frac{1}{2})$ in a single-state SQKD protocol. $U_{F}$ and $U_{R}$ denote the attack operator performed by Eve in the forward  and reverse channel respectively. $U$ is an unitary operator acting on the joint system $\mathcal{H}_{T}\otimes\mathcal{H}_{E}$. We describe the restriction attack strategy as follows:
\begin{enumerate}
\item Alice prepares and sends state $|+\rangle_{T}$ to Bob through the forward channel. Eve intercepts $|+\rangle_{T}$ and resends another state $|e\rangle_{T}$ prepared by herself to Bob, where
\begin{eqnarray}
&& |e\rangle_{T}=\sqrt{\frac{1}{2}+b}|0\rangle_{T}+\sqrt{\frac{1}{2}-b}|1\rangle_{T}.
\end{eqnarray}

\item Bob has two choices when he receives the state $|e\rangle_{T}$.

CTRL : Bob chooses to reflect $|e\rangle_{T}$ back to Alice undisturbed through the reverse channel. Meanwhile, Eve captures the transit state and probes it using unitary operator $U$ acting on the transit state and her own ancilla state. After that Eve resends the transit state to Alice and keeps the ancilla state in her own memory.

SIFT: Bob chooses to discard the state $|e\rangle_{T}$ and send $|0\rangle_{T}$ to Alice instead. Eve can also perform the same attack during the transmission.

\end{enumerate}

The parameter $b$ can specify the amount of noise introduced in the forward channel. It can be observed by the legitimate users.
Eve probes the state by using a unitary operator $U$ to act on $\mathcal{H}_{T}\otimes \mathcal{H}_{E}$ as follows:

\begin{eqnarray}
&& U|0,0\rangle_{TE} = |0,e_{00}\rangle_{TE}+|1,e_{01}\rangle_{TE},\\
&& U|1,0\rangle_{TE} = |0,e_{10}\rangle_{TE}+|1,e_{11}\rangle_{TE}.
\end{eqnarray}

Since $U$ is unitary, we can derive that

\begin{eqnarray}
&& \langle e_{00}|e_{10}\rangle_{E} + \langle e_{01}|e_{11}\rangle_{E} = 0,\\
&& \langle e_{00}|e_{00}\rangle_{E} + \langle e_{01}|e_{01}\rangle_{E} =1,  \\
&& \langle e_{10}|e_{10}\rangle_{E} + \langle e_{11}|e_{11}\rangle_{E} =1.
\end{eqnarray}

In order to illustrate Eve's attack under the circumstance Bob chooses CTRL  and Alice chooses to measure in $X$ basis, we express $|e\rangle$ in $X$ basis as

\begin{eqnarray}
&& |e\rangle_{T} = \frac{\alpha+\beta}{\sqrt{2}} |+\rangle_{T} + \frac{\alpha-\beta}{\sqrt{2}} |-\rangle_{T},\\
&& \alpha = \sqrt{\frac{1}{2}+b} ,  \beta = \sqrt{\frac{1}{2}-b}.
\end{eqnarray}

According to Eqs. (11) and (12), we can get

\begin{eqnarray}
&& U|+,0\rangle_{TE} = |+,f_{+0}\rangle_{TE}+ |-,f_{+1}\rangle_{TE}, \\
&& U|-,0\rangle_{TE} = |+,f_{-0}\rangle_{TE}+ |-,f_{-1}\rangle_{TE},
\end{eqnarray}
where
\begin{eqnarray}
&& |f_{+0}\rangle_{E} = \frac{1}{2}(|e_{00}\rangle_{E}+ |e_{01}\rangle_{E}+|e_{10}\rangle_{E}+|e_{11}\rangle_{E}), \\
&& |f_{+1}\rangle_{E} = \frac{1}{2}(|e_{00}\rangle_{E}- |e_{01}\rangle_{E}+|e_{10}\rangle_{E}-|e_{11}\rangle_{E}), \\
&& |f_{-0}\rangle_{E} = \frac{1}{2}(|e_{00}\rangle_{E}+ |e_{01}\rangle_{E}-|e_{10}\rangle_{E}-|e_{11}\rangle_{E}),\\
&& |f_{-1}\rangle_{E} = \frac{1}{2}(|e_{00}\rangle_{E}- |e_{01}\rangle_{E}-|e_{10}\rangle_{E}+|e_{11}\rangle_{E}).
\end{eqnarray}

Then we can get

\begin{eqnarray}
&& U|e,0\rangle_{TE} = |+,g_{+}\rangle_{TE}+ |-,g_{-}\rangle_{TE},
\end{eqnarray}
where
\begin{eqnarray}
&& |g_{+}\rangle_{E} = \frac{\alpha}{\sqrt{2}}|e_{00}\rangle_{E}+ \frac{\alpha}{\sqrt{2}}|e_{01}\rangle_{E}\\\nonumber
&& \quad \quad \quad + \frac{\beta}{\sqrt{2}}|e_{10}\rangle_{E} +\frac{\beta}{\sqrt{2}}|e_{11}\rangle_{E},\\
&& |g_{-}\rangle_{E} = \frac{\alpha}{\sqrt{2}}|e_{00}\rangle_{E}- \frac{\alpha}{\sqrt{2}}|e_{01}\rangle_{E}\\\nonumber
&& \quad \quad \quad + \frac{\beta}{\sqrt{2}}|e_{10}\rangle_{E}- \frac{\beta}{\sqrt{2}}|e_{11}\rangle_{E}.
\end{eqnarray}

Next, we model one valid iteration of this protocol as follows:

\begin{enumerate}

\item Alice prepares and sends $|+\rangle_{T}$ to Bob through the forward channel:
\begin{eqnarray}
&& \rho^{1}_{T} = |+\rangle\langle+|_{T}.
\end{eqnarray}

\item Eve performs the restricted operation on the transit state
\begin{eqnarray}
&& \rho^{2}_{T} = U_{F}|+\rangle\langle+|_{T}U^{*}_{F} = |e\rangle\langle e|_{T}.
\end{eqnarray}

\item Bob's action:

(1) SIFT:

\begin{eqnarray}
&& \rho^{3}_{BT} = |1\rangle\langle1|_{B}\otimes |0\rangle\langle0|_{T}.
\end{eqnarray}

(2) CTRL:

\begin{eqnarray}
&& \rho^{4}_{BT} = |0\rangle\langle0|_{B}\otimes|e\rangle\langle e|_{T}.
\end{eqnarray}
Because Bob chooses SIFT or CTRL randomly,  $P(\text{SIFT})=P(\text{CTRL})=\frac{1}{2}$. Therefore, the state after Bob's operation is
\begin{eqnarray}
&& \rho^{5}_{BT} = \frac{1}{2}|0\rangle\langle0|_{B}\otimes|e\rangle\langle e|_{T}+\frac{1}{2}|1\rangle\langle1|_{B}\otimes |0\rangle\langle0|_{T}.
\end{eqnarray}
\item Eve's attack in the reverse channel:

(1) SIFT:

\begin{eqnarray}
&& P(|x\rangle)=|x\rangle\langle x|,\\
&& \rho^{6}_{BTE} = |1\rangle\langle1|_{B}\otimes U|0,0\rangle\langle0,0|_{TE}U^{*}\\\nonumber
&& \quad \quad \quad = |1\rangle\langle1|_{B}\otimes P(|0,e_{00}\rangle_{TE}+|1,e_{01}\rangle_{TE}).
\end{eqnarray}

(2) CTRL:

\begin{eqnarray}
&& \rho^{7}_{BTE} = |0\rangle\langle0|_{B}\otimes U|e,0\rangle\langle e,0|_{TE}U^{*}\\\nonumber
&& \quad \quad \quad =|0\rangle\langle0|_{B}\otimes P(|+,g_{+}\rangle_{TE}+ |-,g_{-}\rangle_{TE}).
\end{eqnarray}
Then the mixed state after Eve's attack is
\begin{eqnarray}
&& \rho^{8}_{BTE} =\frac{1}{2}\rho^{6}_{BTE}+\frac{1}{2}\rho^{7}_{BTE}.
\end{eqnarray}
\item Alice measures in $Z$ or $X$ basis randomly:

(1) Measure in $Z$ basis:

\begin{eqnarray}
&& \sigma^{9}_{ABE} = |0\rangle\langle0|_{A}\otimes|1\rangle\langle1|_{B}\otimes \frac{1}{2}|e_{01}\rangle\langle e_{01}|_{E}\\\nonumber
&& \quad \quad \quad +|0\rangle\langle0|_{A}\otimes|0\rangle\langle 0|_{B}\otimes \frac{1}{4}P(|g_{+}\rangle_{E}-|g_{-}\rangle_{E}).
\end{eqnarray}

(2) Measure in $X$ basis:

\begin{eqnarray}
&&　\sigma^{10}_{ABE} =|1\rangle\langle1|_{A}\otimes|0\rangle\langle0|_{B}\otimes\frac{1}{2}|g_{-}\rangle\langle g_{-}|_{E}\\\nonumber
&& \quad \quad \quad　+|1\rangle\langle1|_{A}\otimes|1\rangle\langle1|_{B}\otimes\frac{1}{4}P(|e_{00}\rangle_{E}-|e_{01}\rangle_{E}).
\end{eqnarray}
Note that $\sigma^{9}_{ABE}$ and $\sigma^{10}_{ABE}$ may not be normalized here.
Then the state after Alice's measurement is
\begin{eqnarray}
&&\rho^{11}_{ABE} =\frac{1}{K}[\sigma^{9}_{ABE}+\sigma^{10}_{ABE}],\\
&& K=tr(\sigma^{9}_{ABE}+\sigma^{10}_{ABE}).
\end{eqnarray}
\end{enumerate}

Let $P(i,j)$ denote the probability that the event $A$ and $B$'s raw key bits are  $i$ and $j$, respectively. Then we can get
\begin{eqnarray}
&& P(0,0)= \frac{1}{4K}tr(P(|g_{+}\rangle_{E}-|g_{-}\rangle_{E}))\\\nonumber
&& \quad \quad \quad =\frac{1}{4K}(1-2Re\langle g_{+}|g_{-}\rangle),\\
&& P(0,1)= \frac{1}{2K}tr(|e_{01}\rangle\langle e_{01}|)=\frac{1}{2K}\langle e_{01}|e_{01} \rangle,\\
&& P(1,0)= \frac{1}{2K}tr(|g_{-}\rangle\langle g_{-}|)= \frac{1}{2K}\langle g_{-}| g_{-} \rangle,\\
&& P(1,1)= \frac{1}{4K}tr(P(|e_{00}\rangle_{E}-|e_{01}\rangle_{E}))\\\nonumber
&& \quad \quad \quad =\frac{1}{4K}(1-2Re\langle e_{00}|e_{01}\rangle).
\end{eqnarray}

$P(1,0)$ denotes the probability that $A$'s raw key bit is $1$ and B's raw key bit is $0$. In other words, Alice measures in the $X$ basis and gets the outcome $-$ when Bob chooses to CTRL. This indicates Alice initially sends $|+\rangle$ but getting $|-\rangle$ finally because of the channel noise. We call it the error rate of $X$-type denoted as $e_{X}$. According to the protocol, we can get
\begin{eqnarray}
&& e_{X}=tr_{E}[(|-\rangle\langle-|\otimes I )(U|e,0\rangle\langle e,0|_{TE}U^{*})]\\\nonumber
&& \quad \quad =\langle g_{-}| g_{-} \rangle.
\end{eqnarray}
Similarly, $P(0,1)$ denotes the probability that Alice measures in $Z$ basis and gets the outcome $1$ when Bob chooses to SIFT. We use $e_{Z}$ to denote the error rate of $Z$-type. Then we can get
\begin{eqnarray}
&& e_{Z}=tr_{E}[(|1\rangle\langle 1|\otimes I )(U|0,0\rangle\langle 0,0|_{TE}U^{*})]\\\nonumber
&& \quad \quad =\langle e_{01}|e_{01} \rangle.
\end{eqnarray}
Here $e_{X}$ and $e_{Z}$ are two statistics that can be observed by Alice and Bob in the reconciliation stage.

$P(i,j)$ $(i,j\in\{0,1\})$ is a probability distribution such that
\begin{eqnarray}
&& \sum_{i,j} P(i,j)=1.
\end{eqnarray}
Then we can derive
\begin{eqnarray}
&& K=\frac{1}{4}(1-2Re\langle g_{+}|g_{-}\rangle)+\frac{1}{2}\langle e_{01}|e_{01} \rangle\\\nonumber
&& \quad \quad +\frac{1}{2}\langle g_{-}| g_{-} \rangle+\frac{1}{4}(1-2Re\langle e_{00}|e_{01}\rangle).
\end{eqnarray}

\subsection{Bounding the final key rate}

According to Eq. (9), we can see that we can get a lower bound of the key rate by bounding the von Neumann entropy. Here we also use the expression

\begin{eqnarray}
&& r=\lim_{N\rightarrow\infty}\frac{\ell(N)}{N}\geq \inf(S(B|E)-H(B|A))\\\nonumber
&& \quad \quad \quad \quad \quad \quad \quad \geq \inf(S(B|ME)-H(B|A)),
\end{eqnarray}
which Krawec applied in \cite{12,15} to give the lower bound on the key rate due to the strong subadditivity of von Neumann entropy expressed as
\begin{eqnarray}
&& S(B|E)\geq S(B|ME),
\end{eqnarray}
where $M$ is a new system introduced to form a compound system $ABME$. Then we introduce a new system $M$ modeled by a two dimensional Hilbert space spanned by the orthonormal basis $\{|0\rangle, |1\rangle\}$. We use the operator $|i\rangle\langle i|_{M}, i\in \{0, 1\}$ to record the outcome of performing an $xor$ operation on $A$ and $B$'s raw key bit.
Considering the mixed state of the joint system after one iteration is
\begin{eqnarray}
&& \rho_{ABE}=\frac{1}{K}[|1\rangle\langle1|_{A}\otimes|0\rangle\langle0|_{B}\otimes\frac{1}{2}|g_{-}\rangle\langle g_{-}|_{E}\\\nonumber
&& \quad \quad \quad +|1\rangle\langle1|_{A}\otimes|1\rangle\langle1|_{B}\otimes\frac{1}{4}P(|e_{00}\rangle_{E}-|e_{01}\rangle_{E})\\\nonumber
&& \quad \quad \quad +|0\rangle\langle0|_{A}\otimes|1\rangle\langle1|_{B}\otimes \frac{1}{2}|e_{01}\rangle\langle e_{01}|_{E}\\\nonumber
&& \quad \quad \quad +|0\rangle\langle0|_{A}\otimes|0\rangle\langle 0|_{B}\otimes \frac{1}{4}P(|g_{+}\rangle_{E}-|g_{-}\rangle_{E})],
\end{eqnarray}
then we can get the mixed state of the system $ABME$:

\begin{eqnarray}
&& \rho_{ABME}=  \frac{1}{K}[|1,0\rangle\langle 1,0|_{AB}\otimes |1\rangle\langle 1|_{M} \otimes \frac{1}{2}|g_{-}\rangle\langle g_{-}|_{E}\\\nonumber
&& \quad \quad \quad \quad   +|1,1\rangle\langle 1,1|_{AB}\otimes |0\rangle\langle 0|_{M} \otimes \frac{1}{4}P(|e_{00}\rangle_{E}-|e_{01}\rangle_{E})\\\nonumber
&& \quad \quad \quad \quad   + |0,1\rangle\langle 0,1|_{AB}\otimes |1\rangle\langle 1|_{M} \otimes \frac{1}{2}|e_{01}\rangle\langle e_{01}|_{E}\\\nonumber
&& \quad \quad \quad \quad   + |0,0\rangle\langle 0,0|_{AB} \otimes |0\rangle\langle 0|_{M} \otimes \frac{1}{4}P(|g_{+}\rangle_{E}-|g_{-}\rangle_{E})].
\end{eqnarray}
Tracing out the system $A$, we can get the state $\rho_{BME}$ as

\begin{eqnarray}
&& \rho_{BME}=   \frac{1}{K}[|0,0\rangle\langle 0,0|_{BM}  \otimes \frac{1}{4}P(|g_{+}\rangle_{E}-|g_{-}\rangle_{E})\\\nonumber
&& \quad \quad \quad +|0,1\rangle\langle 0,1|_{BM}  \otimes \frac{1}{2}|g_{-}\rangle\langle g_{-}|_{E}\\\nonumber
&&\quad \quad \quad   +|1,1\rangle\langle 1,1|_{BM} \otimes \frac{1}{2}|e_{01}\rangle\langle e_{01}|_{E}\\\nonumber
&& \quad \quad \quad  +|1,0\rangle\langle 1,0|_{BM} \otimes \frac{1}{4}P(|e_{00}\rangle_{E}-|e_{01}\rangle_{E})].
\end{eqnarray}
Then we get $\rho_{ME}$ as
\begin{eqnarray}
&& \rho_{ME} =tr_{B}(\rho_{BME})\\\nonumber
&& \quad \quad = |0\rangle\langle 0|_{M} \otimes \frac{1}{4K}P(|g_{+}\rangle_{E}-|g_{-}\rangle_{E})\\\nonumber
&& \quad \quad +|0\rangle\langle 0|_{M} \otimes\frac{1}{4K}P(|e_{00}\rangle_{E}-|e_{01}\rangle_{E}) \\\nonumber
&&  \quad\quad + |1\rangle\langle 1|_{M} \otimes (\frac{1}{2K}|g_{-}\rangle\langle g_{-}|_{E}+\frac{1}{2K}|e_{01}\rangle\langle e_{01}|_{E}).
\end{eqnarray}
The mixed states of some certain compound systems have been derived above. Then we compute their von Neumann entropy one by one to bound the final key rate $r$ . Firstly, we compute the von Neumann entropy of system $BME$. In order to compute $S(\rho_{BME})$, we rewrite it as a classical quantum state
\begin{eqnarray}
&& \rho_{BME}= P(0,0)|0,0\rangle\langle 0,0|_{BM} \otimes \rho^{(0,0)}_{E}\\\nonumber
&& \quad \quad \quad +P(1,0)|0,1\rangle\langle 0,1|_{BM}  \otimes \rho^{(1,0)}_{E}\\\nonumber
&&\quad \quad \quad   +P(0,1)|1,1\rangle\langle 1,1|_{BM} \otimes \rho^{(0,1)}_{E}\\\nonumber
&& \quad \quad \quad  +P(1,1)|1,0\rangle\langle 1,0|_{BM} \otimes \rho^{(1,1)}_{E},
\end{eqnarray}
where
\begin{eqnarray}
&& \rho^{(0,0)}_{E}= \frac{P(|g_{+}\rangle_{E}-|g_{-}\rangle_{E})}{tr(P(|g_{+}\rangle_{E}-|g_{-}\rangle_{E}))},\\
&& \rho^{(1,0)}_{E}= \frac{|g_{-}\rangle\langle g_{-}|_{E}}{\langle g_{-}| g_{-}\rangle},\\
&& \rho^{(0,1)}_{E}= \frac{| e_{01}\rangle\langle e_{01}|}{\langle e_{01}| e_{01}\rangle},\\
&& \rho^{(1,1)}_{E}= \frac{P(|e_{00}\rangle_{E}-|e_{01}\rangle_{E})}{tr(P(|e_{00}\rangle_{E}-|e_{01}\rangle_{E}))}.
\end{eqnarray}
According to Eq. (7),  we can figure out $S(\rho_{BME})$ as
\begin{eqnarray}
&&S(\rho_{BME})=H(P(i,j))_{i,j}+\sum_{i,j}P(i,j)S(\rho^{(i,j)}_{E})\\\nonumber
&& \quad\quad\quad\quad\quad \geq H(P(0,0),P(0,1),P(1,0),P(1,1)).
\end{eqnarray}
Note that here we utilize the truth of $S(\rho^{(i,j)}_{E})\geq 0$.
Next, we compute the von Neumann entropy of system $ME$.  At first, we rewrite $\rho_{ME}$ as
\begin{eqnarray}
&& \rho_{ME}=k_{1}|0\rangle\langle 0|_{M}\otimes \rho^{1}_{E}+k_{2}|1\rangle\langle 1|_{M}\otimes \rho^{2}_{E},
\end{eqnarray}
where
\begin{eqnarray}
&& k_{1}=P(0,0)+P(1,1), k_{2}=P(0,1)+P(1,0),\\
&& \rho^{1}_{E}=\frac{P(|g_{+}\rangle_{E}-|g_{-}\rangle_{E})+P(|e_{00}\rangle_{E}-|e_{01}\rangle_{E})}{4(q(0,0)+q(1,1))},\\
&& \rho^{2}_{E}=\frac{|g_{-}\rangle\langle g_{-}|_{E}+|e_{01}\rangle\langle e_{01}|_{E}}{2(q(0,1)+q(1,0))},\\
&& q(0,0)=\frac{1}{4}tr(P(|g_{+}\rangle_{E}-|g_{-}\rangle_{E}))\\\nonumber
&& \quad \quad \quad =\frac{1}{4}(1-2Re\langle g_{+}|g_{-}\rangle),\\
&& q(0,1)=\frac{1}{2}tr(|e_{01}\rangle\langle e_{01}|_{E})=\frac{1}{2}\langle e_{01}|e_{01}\rangle,\\
&& q(1,0)=\frac{1}{2}tr(|g_{-}\rangle\langle g_{-}|_{E})=\frac{1}{2}\langle g_{-}|g_{-}\rangle,\\
&& q(1,1)=\frac{1}{4}tr(P(|e_{00}\rangle_{E}-|e_{01}\rangle_{E}))\\\nonumber
&& \quad\quad\quad=\frac{1}{4}(1-2Re\langle e_{00}|e_{01}\rangle).
\end{eqnarray}
We can see $\rho_{ME}$ is a classical-quantum state. Then $S(\rho_{ME})$ can be figured out as
\begin{eqnarray}
&& S(\rho_{ME})=h(k_{1})+k_{1}S(\rho^{1}_{E})+k_{2}S(\rho^{2}_{E}).
\end{eqnarray}
Therefore, we can find an upper bound of $S(\rho_{ME})$  as
\begin{eqnarray}
&& S(\rho_{ME})\leq h(k_{1})+k_{2}+k_{1}S(\rho^{1}_{E})
\end{eqnarray}
since $\rho^{2}_{E}$ is a two dimensional density operator, satisfying
\begin{eqnarray}
&& S(\rho^{2}_{E})\leq 1.
\end{eqnarray}
Then we can further get the lower bound on the key rate $r$ as
\begin{eqnarray}
  \nonumber  r&\geq& H(P(i,j)_{ij})-h(k_{1})-k_{2}-k_{1}S(\rho^{1}_{E})\\
&&-H(B|A).
\end{eqnarray}
In order to derive an expression of a lower bound of $r$, we need to express $S(\rho^{1}_{E})$ and $H(B|A)$ by using the observable statistics. Then we compute $S(\rho^{1}_{E})$ and $H(B|A)$ one by one.

First of all, we compute $S(\rho^{1}_{E})$.  According to Eqs. (8) and (9), we need to get all eigenvalues of $\rho^{1}_{E}$. Let $|l_{1}\rangle_{E}=|g_{+}\rangle_{E}-|g_{-}\rangle_{E}$, $|l_{2}\rangle_{E}=|e_{00}\rangle_{E}-|e_{01}\rangle_{E}$. Then we can rewrite $\rho^{1}_{E}$ as follows:
\begin{eqnarray}
&& \rho^{1}_{E}=\frac{|l_{1}\rangle\langle l_{1}|_{E}+|l_{2}\rangle\langle l_{2}|_{E}}{\langle l_{1}|l_{1}\rangle+\langle l_{2}|l_{2}\rangle}.
\end{eqnarray}
Let $|l_{1}\rangle= x |\xi\rangle$ and $|l_{2}\rangle= y |\xi\rangle + z |\eta\rangle$, where $x,y,z\in\mathbb{C}$, $\langle\xi|\xi\rangle=\langle\eta|\eta\rangle=1$ and $\langle\xi|\eta\rangle=0$. This indicates:
\begin{eqnarray}
&&|x|^{2}=\langle l_{1}| l_{1}\rangle = 4q(0,0),\\
&&|y|^{2}+|z|^{2}=\langle l_{2}| l_{2}\rangle = 4q(1,1),\\
&& x^{*}y = \langle l_{1}| l_{2}\rangle,\\
&&|y|^{2}=\frac{|\langle l_{1}|l_{2}\rangle|^{2}}{|x|^{2}}.
\end{eqnarray}
Then we can write $\rho^{1}_{E}$ as a matrix in the basis of $\{|\xi\rangle, |\eta\rangle\}$:
\begin{equation}
\rho^{1}_{E}=\frac{1}{|x|^{2}+|y|^{2}+|z|^{2}}
\left(                 
  \begin{array}{cc}   
    |x|^{2}+|y|^{2} & \quad yz^{*} \\  
    \quad y^{*}z \quad\quad  & \quad|z|^{2} \\  
  \end{array}
\right).                 
\end{equation}
Its eigenvalues are
\begin{eqnarray}
&& \lambda_{\pm}=\frac{1}{2}\pm\frac{\sqrt{k_{3}+2k_{4}}}{2(|x|^{2}+|y|^{2}+|z|^{2})},
\end{eqnarray}
where
\begin{eqnarray}
&& k_{3}=|x|^{4}+|y|^{4}+|z|^{4},\\
&& k_{4}=|x|^{2}|y|^{2}+|y|^{2}|z|^{2}-|x|^{2}|z|^{2}.
\end{eqnarray}

Through some mathematical skills and combining Eqs. (77)(78)(79) and (80), we can have
\begin{eqnarray}
&& \lambda_{\pm}=\frac{1}{2}\pm\frac{\sqrt{4(q(0,0)-q(1,1))^{2}+|\langle l_{1}|l_{2}\rangle|^{2}}}{4(q(0,0)+q(1,1))}.
\end{eqnarray}
Thus, we can compute $S(\rho^{1}_{E})$ as
\begin{eqnarray}
&& S(\rho^{1}_{E})=h(\lambda_{+}).
\end{eqnarray}
 From Eq. (81), we can see $\lambda_{+}\geq\frac{1}{2}$, and thus $h(\lambda_{+})$ will increase as $\lambda_{+}$ decreases. Therefore, we can find an upper bound of $S(\rho^{1}_{E})$ by finding a lower bound of  $|\langle l_{1}|l_{2}\rangle|^{2}$. Assume $\mathcal{B}\geq0$ is a lower bound of $|\langle l_{1}|l_{2}\rangle|$ and define
\begin{eqnarray}
&& \lambda=\frac{1}{2}+\frac{\sqrt{4(q(0,0)-q(1,1))^{2}+\mathcal{B}^{2}}}{4(q(0,0)+q(1,1))}.
\end{eqnarray}
Therefore, we have found an upper bound of $S(\rho^{1}_{E})$ as
\begin{eqnarray}
&& S(\rho^{1}_{E})\leq h(\lambda).
\end{eqnarray}

Next, we compute $H(B|A)$ by the observable statistics $P(i,j)_{ij}$. We can easily get
\begin{eqnarray}
&& H(BA)=H(\{P(i,j)\}_{ij}).
\end{eqnarray}
Because
\begin{eqnarray}
&& P_{A}(0)=P(0,0)+P(0,1),\\
&& P_{A}(1)=P(1,0)+P(1,1),
\end{eqnarray}
where $P_{A}(i)$, $i\in\{0,1\}$ means the probability of the event that Alice's raw key bit is $i$.
Then we have,
\begin{eqnarray}
&& H(A)=h(P_{A}(0))=h(P(0,0)+P(0,1)).
\end{eqnarray}
Thus,
\begin{eqnarray}
&&H(B|A)=H(BA)-H(A)\\\nonumber
&&\quad\quad\quad\quad =H(\{P(i,j)\}_{ij})-h(P(0,0)+P(0,1)).
\end{eqnarray}
Therefore, we can obtain a lower bound on the final key rate as
\begin{eqnarray}
r\geq h(P(0,0)+P(0,1))-h(k_{1})-k_{2}-k_{1}h(\lambda).
\end{eqnarray}
From the above inequation, we can see all the parameters can be estimated by $A$ and $B$ except $|\langle l_{1}|l_{2}\rangle|$'s lower bound $\mathcal{B}$. Next, we also consider to use some other observable statistics to determine a value of $\mathcal{B}$.

\subsection{Bounding $|\langle l_{1}|l_{2}\rangle|$ using observable statistics}
In this part, we need to express a lower bound of $|\langle l_{1}|l_{2}\rangle|$ by using some statistics that can be observed by the legitimate users. Recall that $|l_{1}\rangle_{E}=|g_{+}\rangle_{E}-|g_{-}\rangle_{E}$ and $|l_{2}\rangle_{E}=|e_{00}\rangle_{E}-|e_{01}\rangle_{E}$. From Eq. (25) and Eq. (26), we can derive that
\begin{eqnarray}
&& |l_{1}\rangle_{E}=\sqrt{2}\alpha|e_{01}\rangle_{E}+\sqrt{2}\beta|e_{11}\rangle_{E}.
\end{eqnarray}
Thus,
\begin{eqnarray}
&& \langle l_{1}|l_{2}\rangle=\sqrt{2}\alpha\langle e_{01}|e_{00}\rangle-\sqrt{2}\alpha\langle e_{01}|e_{01}\rangle\\\nonumber
&& \quad\quad\quad -\sqrt{2}\beta\langle e_{11}|e_{01}\rangle+\sqrt{2}\beta\langle e_{11}|e_{00}\rangle.
\end{eqnarray}
Then, we can easily get
\begin{eqnarray}
&& \langle l_{2}|l_{1}\rangle=\overline{\langle l_{1}|l_{2}\rangle}=\sqrt{2}\alpha\langle e_{00}|e_{01}\rangle-\sqrt{2}\alpha\langle e_{01}|e_{01}\rangle\\\nonumber
&& \quad\quad\quad \quad\quad\quad \quad -\sqrt{2}\beta\langle e_{01}|e_{11}\rangle+\sqrt{2}\beta\langle e_{00}|e_{11}\rangle.
\end{eqnarray}
Considering that
\begin{eqnarray}
\nonumber  |\langle l_{2}|l_{1}\rangle|&=& |\langle l_{1}|l_{2}\rangle|\\
&=&\sqrt{(Re\langle l_{2}|l_{1}\rangle)^{2}+(Im\langle l_{2}|l_{1}\rangle)^{2}},
\end{eqnarray}
we can specify $\mathcal{B}$ as
\begin{eqnarray}
&& |\langle l_{2}|l_{1}\rangle|\geq\mathcal{B}=\sqrt{(Re\langle l_{2}|l_{1}\rangle)^{2}} =|Re\langle l_{2}|l_{1}\rangle|.
\end{eqnarray}
More specifically,
\begin{eqnarray}
&& \mathcal{B}=|Re(\sqrt{2}\alpha\langle e_{00}|e_{01}\rangle-\sqrt{2}\alpha\langle e_{01}|e_{01}\rangle\\\nonumber
&& \quad\quad-\sqrt{2}\beta\langle e_{01}|e_{11}\rangle+\sqrt{2}\beta\langle e_{00}|e_{11}\rangle)|.
\end{eqnarray}
 We define
\begin{eqnarray}
\mathcal{B}=
\begin{cases}
Re\langle l_{2}|l_{1}\rangle,      & Re\langle l_{2}|l_{1}\rangle\geq 0, \\
-Re\langle l_{2}|l_{1}\rangle,   & Re\langle l_{2}|l_{1}\rangle< 0.
\end{cases}
\end{eqnarray}
Thus, we can use observable statistics to bound $|\langle l_{1}|l_{2}\rangle|$ by specifying $Re\langle l_{2}|l_{1}\rangle$. In order to specify $Re\langle l_{2}|l_{1}\rangle$, we need to specify $\sqrt{2}\alpha Re\langle e_{00}|e_{01}\rangle$, $\sqrt{2}\alpha Re\langle e_{01}|e_{01}\rangle$, $\sqrt{2}\beta Re\langle e_{01}|e_{11}\rangle$ and $\sqrt{2}\beta Re \langle e_{00}|e_{11}\rangle$. Next, we specify them one by one.

\begin{enumerate}
\item $\sqrt{2}\alpha Re\langle e_{00}|e_{01}\rangle$:

From Eq. (67), we can get
\begin{eqnarray}
&&  \sqrt{2}\alpha Re\langle e_{00}|e_{01}\rangle=\frac{\sqrt{2}\alpha}{2}-2\sqrt{2}\alpha q(1,1).
\end{eqnarray}

\item $\sqrt{2}\alpha Re\langle e_{01}|e_{01}\rangle$:

According to Eq. (45), we can have
\begin{eqnarray}
&&  \langle e_{01}|e_{01}\rangle=e_{Z}.
\end{eqnarray}
This implies $\langle e_{01}|e_{01}\rangle$ is a real number, and therefore, $Re\langle e_{01}|e_{01}\rangle=\langle e_{01}|e_{01}\rangle$. Then
we can specify it as
\begin{eqnarray}
&& \sqrt{2}\alpha Re\langle e_{01}|e_{01}\rangle=\sqrt{2}\alpha \langle e_{01}|e_{01}\rangle=\sqrt{2}\alpha e_{Z}.
\end{eqnarray}

\item $\sqrt{2}\beta Re \langle e_{01}|e_{11}\rangle$:

At this point, we focus on the process that Bob chooses CTRL and Alice measures in the $Z$-basis and observes $|1\rangle$. We use $P(K_{A}=0|K_{B}=0)$ to denote the probability of the event that Alice measures in $Z$-basis and observes $|1\rangle$ under the circumstance that Bob chooses to CTRL. We abbreviate $P(K_{A}=0|K_{B}=0)$ as $P(0|0)$. Firstly, we model this process as
\begin{eqnarray}
&& \quad \rho_{TE}=U|e,0\rangle\langle e,0|_{TE}U^{*}\\\nonumber
&& \quad\quad\quad =P(|0,\alpha e_{00}+\beta e_{10}\rangle_{TE}+|1,\alpha e_{01}+\beta e_{11}\rangle_{TE}).
\end{eqnarray}
Then Alice measures in $Z$-basis and observes $|1\rangle$ with the probability $P(0|0)$.
\begin{eqnarray}
&& \quad P(0|0)=tr(|1\rangle\langle 1|_{T}\otimes I)\rho_{TE}\\\nonumber
&& \quad\quad\quad\quad =2\alpha\beta Re\langle e_{01}|e_{11}\rangle+\alpha^{2}\langle e_{01}|e_{01}\rangle+\beta^{2}\langle e_{11}|e_{11}\rangle\\\nonumber
&& \quad\quad\quad\quad =2\alpha\beta Re\langle e_{01}|e_{11}\rangle+(\alpha^{2}-\beta^{2})\langle e_{01}|e_{01}\rangle+\beta^{2}.
\end{eqnarray}
Here we have used Eqs.(18)(19) and the assumption of symmetrical property which is often used in QKD security proof.
 Thus, we can specify $\sqrt{2}\beta Re \langle e_{01}|e_{11}\rangle$ as
\begin{eqnarray}
&& \quad\quad\quad\sqrt{2}\beta Re \langle e_{01}|e_{11}\rangle\\\nonumber
&&\quad\quad\quad =\frac{\sqrt{2}}{2\alpha}[P(0|0)-(\alpha^{2}-\beta^{2})e_{Z}-\beta^{2}].
\end{eqnarray}
\item $\sqrt{2}\beta Re \langle e_{00}|e_{11}\rangle$:

At this time, we pay attention to the process that Alice measures in the $X$-basis and observes $|-\rangle$ when Bob chooses to CTRL. We use $P(1|0)$ to denote the probability of the event that Alice measures in the $X$-basis and observes $|-\rangle$ under the circumstance Bob chooses to CTRL. Then we can compute it as
\begin{eqnarray}
&& P(1|0)=\langle g_{-}|g_{-}\rangle\\\nonumber
&& \quad \quad \quad  =\frac{1}{2}-\alpha^{2}Re\langle e_{00}|e_{01}\rangle-\alpha\beta Re\langle e_{00}|e_{11}\rangle\\\nonumber
&& \quad\quad\quad -\alpha\beta Re\langle e_{01}|e_{10}\rangle-\beta^{2}Re\langle e_{10}|e_{11}\rangle.
\end{eqnarray}
Then we can derive that
\begin{eqnarray}
&& \sqrt{2}\beta Re \langle e_{00}|e_{11}\rangle=\frac{\sqrt{2}}{\alpha}[\frac{1}{2}-P(1|0) -\alpha^{2}(\frac{1}{2}\\\nonumber
&& -2q(1,1))-\alpha\beta Re\langle e_{01}|e_{10}\rangle-\beta^{2}Re\langle e_{10}|e_{11}\rangle].
\end{eqnarray}
We can see the right side of the Eq. (105) still contains the expression $Re\langle e_{01}|e_{10}\rangle$ and $Re\langle e_{10}|e_{11}\rangle$. Here we cannot specify them using the observable statistics, but we can bound them by using the Cauchy-Schwarz inequality:
\begin{eqnarray}
&& |\langle e_{01}|e_{10}\rangle|\leq\sqrt{\langle e_{01}|e_{01}\rangle\langle e_{10}|e_{10}\rangle}\\\nonumber
&& \quad\quad\quad\quad\quad=|\langle e_{01}|e_{01}\rangle|=e_{Z},\\
&& |\langle e_{10}|e_{11}\rangle|\leq\sqrt{\langle e_{10}|e_{10}\rangle\langle e_{11}|e_{11}\rangle}\\\nonumber
&& \quad\quad\quad\quad\quad =\sqrt{e_{Z}(1-e_{Z})}.
\end{eqnarray}
Thus,
\begin{eqnarray}
&& Re\langle e_{01}|e_{10}\rangle\leq\sqrt{(Re\langle e_{01}|e_{10}\rangle)^{2}}\\\nonumber
&& \quad\quad\quad\quad\quad\leq\sqrt{\langle e_{01}|e_{01}\rangle\langle e_{10}|e_{10}\rangle}=e_{Z},\\
&& Re\langle e_{10}|e_{11}\rangle\leq\sqrt{(Re\langle e_{10}|e_{11}\rangle)^{2}}\\\nonumber
&& \quad\quad\quad\quad\quad\leq\sqrt{\langle e_{10}|e_{10}\rangle\langle e_{11}|e_{11}\rangle}=\sqrt{e_{Z}(1-e_{Z})}.
\end{eqnarray}
Then we can find a lower bound of $\sqrt{2}\beta Re \langle e_{00}|e_{11}\rangle$ as
\begin{eqnarray}
&& \sqrt{2}\beta Re \langle e_{00}|e_{11}\rangle\geq \frac{\sqrt{2}}{\alpha}[\frac{1}{2}-P(1|0)-\alpha^{2}(\frac{1}{2}\\\nonumber
&& \quad \quad \quad \quad \quad  -2q(1,1))-\alpha\beta e_{Z}-\beta^{2}\sqrt{e_{Z}(1-e_{Z})}].
\end{eqnarray}

\end{enumerate}
From the above, we can get a lower bound on $|\langle l_{1}|l_{2}\rangle|$ as
\begin{eqnarray}
&& |\langle l_{1}|l_{2}\rangle|\geq \mathcal{B}= \frac{\sqrt{2}\alpha}{2}-2\sqrt{2}\alpha q(1,1)-\sqrt{2}\alpha e_{Z}\\\nonumber
&& \quad \quad \quad \quad \quad-\frac{\sqrt{2}}{2\alpha}[P(0|0)-(\alpha^{2}-\beta^{2})e_{Z}-\beta^{2}]\\\nonumber
&& \quad \quad \quad \quad\quad +\frac{\sqrt{2}}{\alpha}[\frac{1}{2}-P(1|0)-\alpha^{2}(\frac{1}{2}-2q(1,1))\\\nonumber
&& \quad \quad \quad \quad\quad -\alpha\beta e_{Z}-\beta^{2}\sqrt{e_{Z}(1-e_{Z})}].
\end{eqnarray}
Note that here we can ensure $\mathcal{B}$ to be positive by controlling the noise in the forward and reverse quantum channel. This is reasonable because the protocol should be aborted if there is too much noise.

From the above, all the parameters appeared in the right hand of Eq. (90) are specified by the observable statistics. Then we have found a lower bound of the key rate $r$ which is expressed as a function of channel parameters because all the observable statistics are determined by the quantum channel. Thus, we can compute a threshold value of the error rate such that the key rate $r$ can always be positive when all the errors are less than this value. In other words, the secure key can be established successfully as long as all the error rates are less than the threshold value. Finally, the full security proof restricted on Eve's collective attack is completed.

In order to get the whole unconditional security proof, we need to spread the circumstance of collective attack to general attack. Fortunately, Renner et al \cite{17}  proved that it suffices to consider the collective attack if protocols are permutation invariant. Next, we will show our protocol is permutation invariant by reducing it to a $B92$ protocol with small modifications. Though our protocol relies on a two-way quantum channel, Krawec \cite{14} has proved that all the attacks can be equivalent to a restricted attack. Then we can reduce our protocol to a fully quantum key distribution protocol with one-way quantum channel. Specifically, it can be reduced to a protocol that Bob prepares a state of set $\{|0\rangle, |e\rangle\}$ at random and Alice measures in $Z$ or $X$ basis randomly, which is a kind of modified $B92$ protocol. Renner et al \cite{17} showed that $B92$ is permutation invariant. Therefore, our protocol is permutation invariant as well. Thus we can derive that our protocol can be secure against general attack. The whole unconditional security proof is completed.

\subsection{Example}
In this part, we illustrate how to compute the threshold value of the error rates under the circumstance that the reverse channel is a depolarizing one with parameter $p$. The depolarization channel is a typical scenario considered in the unconditional security proofs of some other protocols \cite{12,19,20}. It can be specified as follows:

\begin{eqnarray}
&& \xi_{p}(\rho)=(1-p)\rho+\frac{p}{2}I,
\end{eqnarray}
where $I$ is the identity operator.

We model Eve's attack  in the reverse channel after Bob's action as follows:
\begin{enumerate}
\item SIFT:
\begin{eqnarray}
&& \xi_{p}(|0\rangle\langle 0|_{T})=(1-\frac{p}{2})|0\rangle\langle 0|_{T}+\frac{p}{2}|1\rangle\langle1|_{T}.
\end{eqnarray}

\item CTRL:
\begin{eqnarray}
&& \xi_{p}(|e\rangle\langle e|_{T})=(1-\frac{p}{2})|e\rangle\langle e|_{T}+\frac{p}{2}|e^{\bot}\rangle\langle e^{\bot}|_{T},
\end{eqnarray}
where $|e^{\bot}\rangle$ is a state orthogonal to $|e\rangle$, that is to say,
\begin{eqnarray}
|e^{\bot}\rangle=\sqrt{\frac{1}{2}-b}|0\rangle-\sqrt{\frac{1}{2}+b}|1\rangle.
\end{eqnarray}
\end{enumerate}
Then we can get the mixed state of the compound system after an iteration:
\begin{eqnarray}
\nonumber \rho=\frac{1}{2}|1\rangle\langle 1|_{B}\otimes\xi_{p}(|0\rangle\langle 0|_{T})\\ +\frac{1}{2}|0\rangle\langle 0|_{B}\otimes\xi_{p}(|e\rangle\langle e|).
\end{eqnarray}
Next, we compute the parameters appeared in Eqs.(94) and (115) one by one.

Firstly, we compute $q(i, j), i,j \in \{0,1\}$ in terms of the parameters of $b$ and $p$ :
\begin{eqnarray}
&& q(0, 0)=tr[(|0\rangle\langle 0|_{B}\otimes|1\rangle\langle 1|_{T})\rho]\\\nonumber
&&\quad\quad\quad  =\frac{1}{4}-\frac{b}{2}+\frac{pb}{2},\\
&& q(0, 1)=tr[(|1\rangle\langle 1|_{B}\otimes|1\rangle\langle 1|_{T})\rho]=\frac{p}{4},\\
&& q(1, 0)=tr[(|0\rangle\langle 0|_{B}\otimes|-\rangle\langle -|_{T})\rho]\\\nonumber
&& \quad\quad\quad =\frac{1}{4}-\frac{(1-p)\sqrt{1-4b^{2}}}{4},\\
&& q(1, 1)=tr[(|1\rangle\langle 1|_{B}\otimes|-\rangle\langle -|_{T})\rho]=\frac{1}{4}.
\end{eqnarray}

Then we can derive
\begin{eqnarray}
&& K=\sum_{i,j}q(i, j)\\\nonumber
&& \quad =\frac{3}{4}-\frac{b}{2}+\frac{pb}{2}+\frac{p}{4}-\frac{(1-p)\sqrt{1-4b^{2}}}{4}.
\end{eqnarray}
Thus we can get $P(i, j), i,j\in \{0, 1\}$:
\begin{eqnarray}
&& P(0, 0)=\frac{1-2b+2pb}{3-2b+2pb+p-(1-p)\sqrt{1-4b^{2}}},\\
&& P(0, 1)=\frac{p}{3-2b+2pb+p-(1-p)\sqrt{1-4b^{2}}},\\
&& P(1, 0)=\frac{1-(1-p)\sqrt{1-4b^{2}}}{3-2b+2pb+p-(1-p)\sqrt{1-4b^{2}}},\\
&& P(1, 1)=\frac{1}{3-2b+2pb+p-(1-p)\sqrt{1-4b^{2}}}.
\end{eqnarray}
Next, we compute $e_{Z}$, $P(0|0)$ and $P(1|0)$ as follows:
\begin{eqnarray}
&& e_{Z}=tr[|1\rangle\langle 1|\xi_{p}(|0\rangle\langle 0|)]=\frac{p}{2};\\
&& P(0|0)=tr[|1\rangle\langle 1|\xi_{p}(|e\rangle\langle e|)]=\frac{1}{2}-b+pb;\\
&& P(1|0)=tr[|-\rangle\langle -|\xi_{p}(|e\rangle\langle e|)]\\\nonumber
&& \quad\quad \quad =\frac{1}{2}-\sqrt{\frac{1}{4}-b^{2}}+p\sqrt{\frac{1}{4}-b^{2}}.
\end{eqnarray}
Thus, we can get a lower bound on the key rate $r$ according to Eq. (90) as
\begin{eqnarray}
&& r\geq f(b,p),\\
&& f(b,p)=h(P(0,0)+P(0,1))-h(P(0,0)\\\nonumber
&& +P(1,1))-(P(0,1)+P(1,0))-(P(0,0)+P(1,1))h(\lambda).
\end{eqnarray}
Then we can specify $f(b,p)$ as
\begin{eqnarray}
&& f(b,p)=h(\frac{1-2b+2pb+p}{K'})-h(\frac{2-2b+2pb}{K'}) \\\nonumber
&& \quad \quad \quad -\frac{1+p-(1-p)\sqrt{1-4b^{2}}}{K'}-\frac{2-2b+2pb}{K'}h(\lambda),\\
&& K'=3+p+(p-1)\sqrt{1-4b^{2}}+2b(p-1),\\
&& \lambda=\frac{1}{2}+\frac{\sqrt{(pb-b)^{2}+\mathcal{B}^{2}}}{2-2b+2pb},\\
&& \mathcal{B}= \frac{2}{\sqrt{1+2b}}[\sqrt{1-4b^{2}}(\frac{1}{2}-\frac{3p}{4})\\\nonumber
&& \quad\quad -\frac{1}{2}(\frac{1}{2}-b)\sqrt{2p-p^{2}}]-\frac{p}{2}\sqrt{1-2p}.
\end{eqnarray}

A graph of the lower bound of the key rate $r$ as a function of $p$ for different $b$ is shown in Figure 1. In the graph, we can see when $b=0$, the key rate $r$ is positive for all $p\leq 0.0692$, which means that when $e_{Z}=\frac{p}{2}\leq 3.46\%$, the key rate will always be positive. Different values of $b$ correspond to different threshold values which assure the key rate $r$ is positive. We can see when the absolute value of $b$ is far from $0$, the threshold value becomes smaller, which demonstrates that the noise in the forward channel has an effect on the final key rate in some extent.

\begin{figure}
\center
\includegraphics[height=6cm, width=8cm]{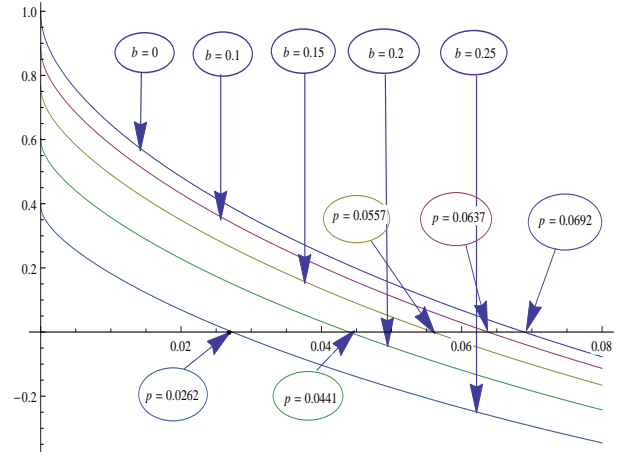}
\caption{\label{figure1}  A graph of our lower bound on the key rate of this SQKD protocol as a function of the depolarization channel parameter $p$ for different $b$. Note that, $e_{Z}=\frac{p}{2}$.}
\end{figure}

A graph of the lower bound of the key rate $r$ as a function of $b$ for different $p$ is shown in Figure 2. In this graph, we can see the lower bound decreases sharply when the parameter $p$ increases a little. This indicates that the noise in the reverse channel affect it more evident. Therefore, we have to make more efforts to control the noise in the reverse channel when the protocol is implemented.

\begin{figure}
\center
\includegraphics[height=6cm, width=8cm]{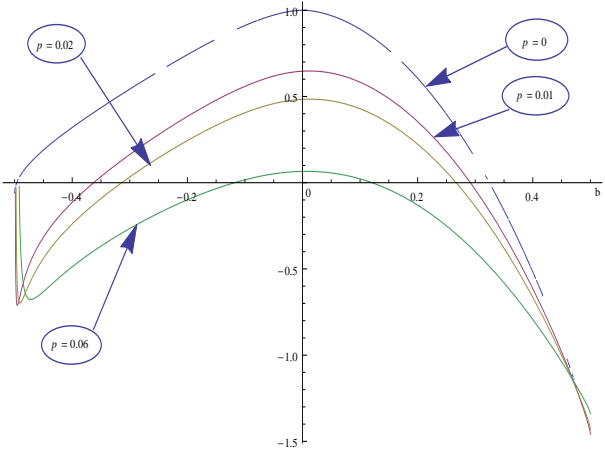}
\caption{\label{figure1}  A graph of the lower bound on the key rate of this SQKD protocol as a function of the forward channel parameter $b$ for different $p$.}
\end{figure}

Next, we make a comparison of the protocol with Krawec's protocol in \cite{12}.  It is proved that Krawec's single-state protocol can endure the maximum bit error rate $e_{Z}=\frac{p}{2}\leq 5.36\%$ when the forward channel parameter $b=0$ \cite{12}. A graph of the lower bound of the key rate $r$ as a function of $p$ in case of $b=0$ of the two compared protocols is shown in Figure 3. In this graph, we can see $p_{1}=0.1072>p_{2}=0.0692$ and the lower bound's decreasing speed is $2>1$. These indicate that our protocol can endure less noise under the circumstance of $b=0$. Under this circumstance, Krawec's protocol can be considered as a modified three-state BB84 protocol. Specifically, the sender Bob prepares one of state from the set $\{|0\rangle, |1\rangle,|+\rangle \}$ each iteration, but they drops all the iterations when he sends $|1\rangle$ to Alice after quantum communication. It is proved that the asymmetric three-state BB84 can tolerate $e_{Z}<4.36\%$ quantum bit error rate \cite{21}, comparing to $e_{Z}<9.81\%$ of the symmetric three-state BB84 \cite{21,22}. Similarly, our protocol can be seen as the B92 protocol mentioned previously. Exactly, Bob prepares a state from the set $\{|0\rangle, |+\rangle \}$ each iteration. In Ref. \cite{23}, it is proved that B92 can tolerate depolarizing rate $p'<0.034$($e_{Z}=\frac{2p'}{3}<2.27\%$). Then the depolarizing rate has been improved to $p'<0.036$($e_{Z}=\frac{2p'}{3}<2.4\%$) in \cite{24}. Finally, Ryutaroh Matsumoto improved the depolarizing rate to $p'<0.065$($e_{Z}=\frac{2p'}{3}<4.33\%$) through convex optimization method \cite{25}. From above, we can see the two SQKD protocols are as secure as their fully counterparts. Though our protocol can tolerate less noise, it can be easily implemented in the real world. This coincides the case in fully quantum cryptography. As we know, B92 is more simple to implement than BB84, it can endure a maximum bit error rate of less than $4.33\%$ comparing to $11\%$ of BB84 protocol \cite{21,24}. In Ref. \cite{21}, it is also shown that more simple the QKD protocol is, less noise it can endure.

\begin{figure}
\center
\includegraphics[height=6cm, width=8cm]{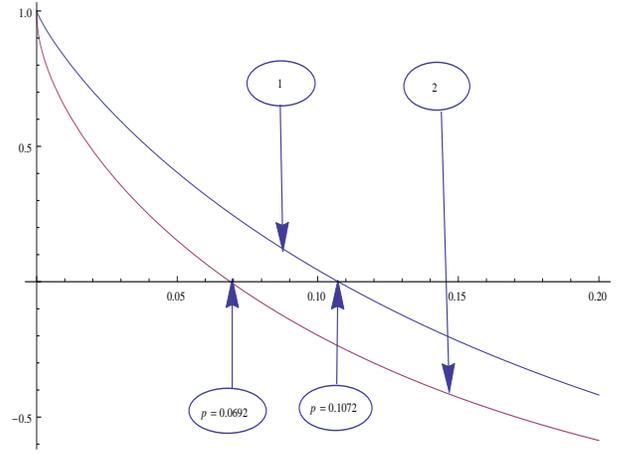}
\caption{\label{figure1}  A graph of the lower bound on the key rate of the two compared protocols as a function of the reverse channel parameter $p$ in case of the forward channel parameter $b=0$. Note that: 1 Krawec's protocol; 2 Our protocol}
\end{figure}

\section{Conclusion}

In this paper, we introduce the idea of B92 into semi-quantum key distribution and design a new SQKD protocol with one qubit. To our best of knowledge, this is the first semi-quantum version of B92 protocol, comparing to BKM07 as the semi-quantum version of BB84. Then we show that it is not only more efficient but also more simplified to implement. Meanwhile, it is demonstrated that our protocol is as secure as some existed SQKD and QKD protocols. We provide an unconditional security proof of our protocol by computing a lower bound of the final key rate in the asymptotic scenario and found a threshold value of errors such that if all the errors are less than this value, the secure key can be established definitely. We show that our scheme can tolerate a maximum bit error rate of $3.46\%$ under the circumstance that there is no noise in the forward channel. It is comparable to  the SQKD protocol BKM07 which can tolerate up to $5.34\%$ error rate under the circumstance that the error rate in $Z$-type is equal to the $X$-type in both forward and reverse quantum channel \cite{15}. It is also comparable to Krawec's newly single-state protocol which can withstand up to error rate of  $5.36\%$ \cite{12}. Though our protocol can endure less noise, it needs fewer quantum resource and equipments which makes it to be more practical and realizable. It has great advantages in practice under the circumstance that the quantum channel is less noisy.

From above, we can see the maximum value of noise that our protocol can tolerate is a little smaller than those of BKM07 and single-state protocol in Ref. \cite{12}. Probably the lower bound of the key rate is not tight here, and we would further improve it to enhance our maximum tolerated value in the future. Maybe Ryutaroh Matsumoto's method in \cite{25} can give us some tips in this direction. More importantly, we talk our unconditional security only in the perfect qubit scenario. It is a challenge problem to consider the unperfect scenario.

\begin{acknowledgements}
The authors would like to thank the referees for their very helpful suggestions that greatly
helped to improve the quality of this paper. The authors thank Xiangfu Zou for checking the protocol designed in the paper and giving useful suggestions. The authors also thank Zhiming Huang for his help in drafting the graph and mathematical software installation. This work is supported in part by the National
Natural Science Foundation of China (Nos. 61272058, 61572532), the Natural Science Foundation of Qiannan Normal
College for Nationalities joint Guizhou Province of China (No. Qian-Ke-He LH Zi[2015]7719), the Natural Science Foundation of Central Government Special Fund for Universities of West China (No. 2014ZCSX17).

\end{acknowledgements}

\bibliography{basename of .bib file}

\end{document}